\pdfoutput=1
\documentclass[12pt,a4paper]{article}
\usepackage{ifthen}
\usepackage{booktabs}
\usepackage{overpic}
\usepackage{relsize}
\usepackage{rotating}
\usepackage[top=1in, bottom=1.25in, left=1in, right=1in]{geometry}

\newboolean{pdflatex}
\setboolean{pdflatex}{true}

\newboolean{articletitles}
\setboolean{articletitles}{true}

\newboolean{uprightparticles}
\setboolean{uprightparticles}{false}

\newboolean{inbibliography}
\setboolean{inbibliography}{false}

\usepackage{microtype}
\usepackage{lineno}
\usepackage{xspace}
\usepackage{caption}

\usepackage{graphicx}
\usepackage{color}
\usepackage{colortbl}
\graphicspath{{./figs/}}

\usepackage{amsmath}
\usepackage{amssymb}
\usepackage{amsfonts}
\usepackage{upgreek}

\newcommand*\patchAmsMathEnvironmentForLineno[1]{%
\expandafter\let\csname old#1\expandafter\endcsname\csname #1\endcsname
\expandafter\let\csname oldend#1\expandafter\endcsname\csname
end#1\endcsname
 \renewenvironment{#1}%
   {\linenomath\csname old#1\endcsname}%
   {\csname oldend#1\endcsname\endlinenomath}%
}
\newcommand*\patchBothAmsMathEnvironmentsForLineno[1]{%
  \patchAmsMathEnvironmentForLineno{#1}%
  \patchAmsMathEnvironmentForLineno{#1*}%
}
\AtBeginDocument{%
\patchBothAmsMathEnvironmentsForLineno{equation}%
\patchBothAmsMathEnvironmentsForLineno{align}%
\patchBothAmsMathEnvironmentsForLineno{flalign}%
\patchBothAmsMathEnvironmentsForLineno{alignat}%
\patchBothAmsMathEnvironmentsForLineno{gather}%
\patchBothAmsMathEnvironmentsForLineno{multline}%
\patchBothAmsMathEnvironmentsForLineno{eqnarray}%
}

\usepackage{hyperref}
\usepackage[all]{hypcap}

 \mathchardef\PLambda="7103
\def\Lz          {{\ensuremath{\PLambda}}\xspace}
\def\Kz      {{\ensuremath{K^0}}\xspace}
\def\pt         {\mbox{$p_{\rm T}$}\xspace}
\newcommand{\tev}{\ifthenelse{\boolean{inbibliography}}{\ensuremath{~T\kern -0.05em eV}\xspace}{\ensuremath{\mathrm{\,Te\kern -0.1em V}}}\xspace}
\newcommand{\gev}{\ensuremath{\mathrm{\,Ge\kern -0.1em V}}\xspace}
\newcommand{\mev}{\ensuremath{\mathrm{\,Me\kern -0.1em V}}\xspace}
\newcommand{\kev}{\ensuremath{\mathrm{\,ke\kern -0.1em V}}\xspace}

\def\ps   {\ensuremath{\mbox{\,ps}}\xspace}
\def\invfb   {\ensuremath{\mbox{\,fb}^{-1}}\xspace}
\def\lhcb {\mbox{LHCb}\xspace}

\def\pythia     {\mbox{\textsc{Pythia}}~8\xspace}
\def\evtgen     {\mbox{\textsc{EvtGen}}\xspace}
\def\photos     {\mbox{\textsc{Photos}}\xspace}
\def\geant      {\mbox{\textsc{Geant4}}\xspace}

\def\Dbar    {{\kern 0.2em\overline{\kern -0.2em D}{}}\xspace}

\def\eg      {\mbox{\itshape e.g.}\xspace}

\usepackage{cite}
\usepackage{mciteplus}

\usepackage{longtable}
\usepackage{chngpage}

\def\jpsi     {{\ensuremath{{J\mskip -3mu/\mskip -2mu\psi\mskip 2mu}}}\xspace}

\def\ptj{\ensuremath{\pt({\rm jet})}\xspace}
\def\z{\ensuremath{z(\jpsi)}\xspace}

\def\tz{\ensuremath{\tilde{t}}\xspace}

\begin{document}

\renewcommand{\thefootnote}{\fnsymbol{footnote}}
\setcounter{footnote}{1}

\begin{titlepage}
\pagenumbering{roman}

% Header ---------------------------------------------------
\vspace*{-1.5cm}
\centerline{\large EUROPEAN ORGANIZATION FOR NUCLEAR RESEARCH (CERN)}
\vspace*{1.5cm}
\hspace*{-0.5cm}
\begin{tabular*}{\linewidth}{lc@{\extracolsep{\fill}}r}
\ifthenelse{\boolean{pdflatex}}
{\vspace*{-2.7cm}\mbox{\!\!\!\includegraphics[width=.14\textwidth]{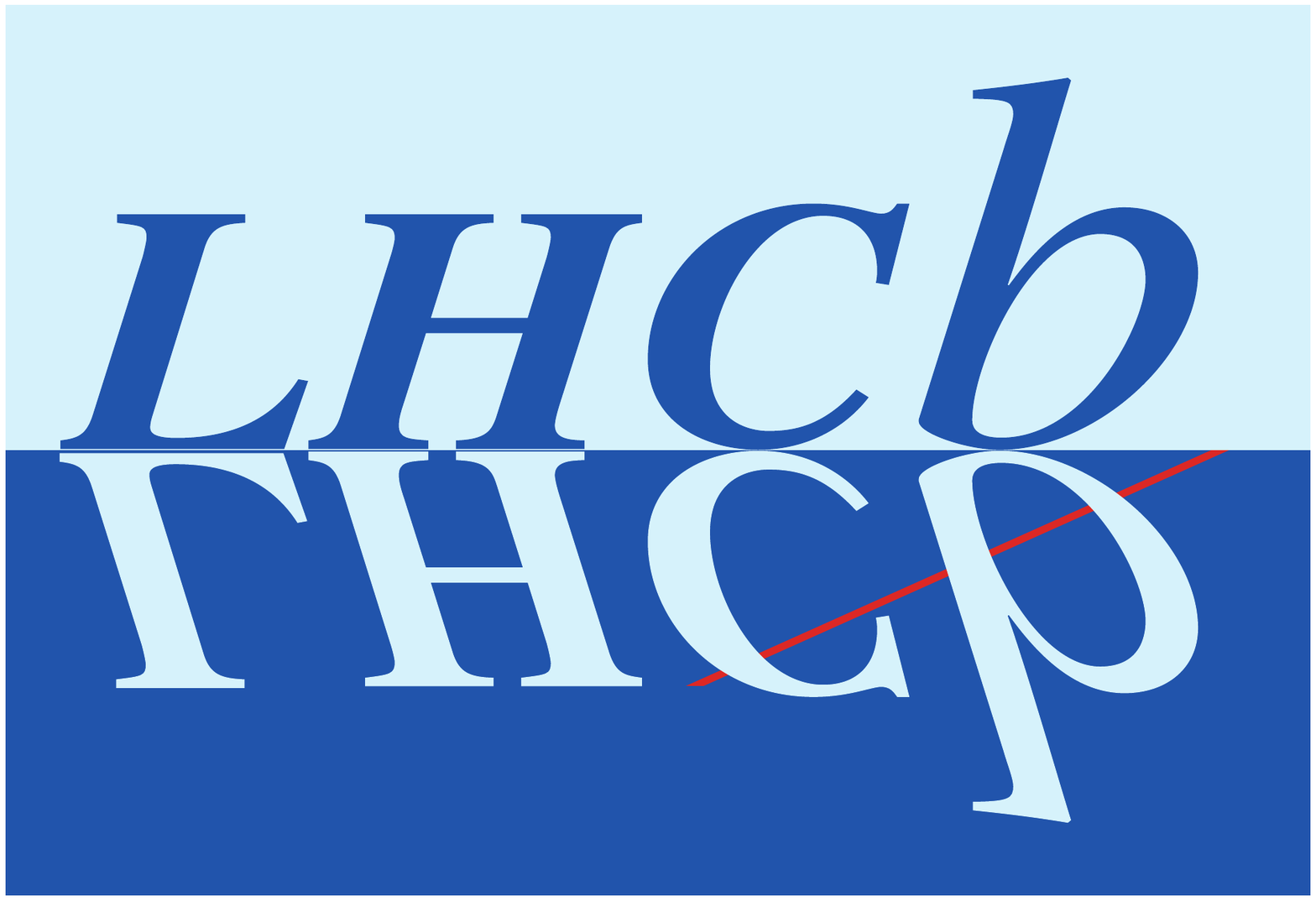}} & &}%
{\vspace*{-1.2cm}\mbox{\!\!\!\includegraphics[width=.12\textwidth]{lhcb-logo.eps}} & &}%
\\
 & & CERN-EP-2017-006 \\
 & & LHCb-PAPER-2016-064 \\
 & & March 24, 2017 \\
 & & \\
\end{tabular*}

\vspace*{4.0cm}

% Title --------------------------------------------------
{\bf\boldmath\huge
\begin{center}
  Study of \jpsi production in jets
\end{center}
}

\vspace*{0.1cm}

% Authors -------------------------------------------------
\begin{center}
The LHCb collaboration\footnote{Authors are listed at the end of this Letter.}
\end{center}

\vspace{\fill}

% Abstract -----------------------------------------------
\begin{abstract}
  \noindent
The production of \jpsi mesons in jets is studied in the forward region of proton-proton collisions
using data collected with the LHCb detector at a center-of-mass energy of 13\tev.
The fraction of the jet transverse momentum carried by the \jpsi meson, $\z\equiv\pt(\jpsi)/\pt({\rm jet})$, is measured
using jets with $\pt({\rm jet}) > 20\gev$ in the pseudorapidity range $2.5 < \eta({\rm jet}) < 4.0$.
The observed \z distribution for \jpsi mesons produced in $b$-hadron decays is consistent with expectations. However, the results for prompt \jpsi production do not agree with predictions based on fixed-order non-relativistic QCD.
This is the first measurement of
the \pt fraction carried by prompt \jpsi mesons in jets
at any experiment.
\end{abstract}

\vspace*{0.5cm}

\begin{center}
  Published in Physical Review Letters
\end{center}

\vspace{\fill}

{\footnotesize
\centerline{\copyright~CERN on behalf of the \lhcb collaboration, license \href{http://creativecommons.org/licenses/by/4.0/}{CC-BY-4.0}.}}
\vspace*{2mm}

\end{titlepage}

\newpage
\setcounter{page}{2}
\mbox{~}
\newpage

\renewcommand{\thefootnote}{\arabic{footnote}}
\setcounter{footnote}{0}

\pagestyle{plain}
\setcounter{page}{1}
\pagenumbering{arabic}

%\linenumbers

\clearpage

The production of \jpsi mesons in hadron-hadron collisions occurs at the transition between the perturbative and non-perturbative regimes of quantum chromodynamics (QCD), resulting in a rich phenomenology that is yet to be fully understood.
Differential \jpsi production cross sections measured at both the Tevatron~\cite{Acosta:2004yw,Abachi:1996jq} and the LHC~\cite{LHCb-PAPER-2015-037,LHCb-PAPER-2013-016,LHCb-PAPER-2012-039,LHCb-PAPER-2011-003,Aad:2015duc,Khachatryan:2015rra,Abelev:2012kr} can be described using the non-relativistic QCD (NRQCD)~\cite{Bodwin:1994jh,Cho:1995vh,Cho:1995ce} effective field theory approach. However, many NRQCD-based calculations~\cite{Campbell:2007ws,Lansberg:2008gk,Gong:2008sn} predict a large degree of transverse polarization, whereas minimal polarization is observed in data~\cite{LHCb-PAPER-2013-008,Abelev:2011md,Abulencia:2007us,Chatrchyan:2013cla}.
This discrepancy indicates that further studies are needed to gain a better understanding of  \jpsi production.

Quarkonium production is often used as a
probe of QCD phenomenology~\cite{Andronic:2015wma}.
In proton-lead ($p$Pb) collisions, \jpsi production is used to study cold-nuclear-matter effects such as parton shadowing and nuclear absorption~\cite{LHCb-PAPER-2013-052,LHCb-PAPER-2015-058,Aad:2015ddl},
while hadron melting in the quark-gluon plasma is investigated using \jpsi production in PbPb collisions~\cite{Aad:2010aa,Khachatryan:2014bva,Sirunyan:2016znt}.
Double-\jpsi production is used to measure the effective cross section for double parton scattering~\cite{Kom:2011nu,Lansberg:2014swa,LHCb-PAPER-2016-057,Abazov:2014qba,Khachatryan:2014iia}, which is commonly assumed to be universal for all processes.
If the prevailing picture of \jpsi meson production directly in parton-parton scattering is not valid, then many quarkonium-production results may need to be reinterpreted.

Another striking, yet untested, prediction of the direct-production paradigm is that \jpsi mesons are largely produced isolated, except for any soft gluonic radiation emitted by the $c\bar{c}$ state
and potentially some particles from the underlying hadron-hadron collision.
An alternative to the standard approach, which is also based on NRQCD,
is the calculation of \jpsi meson production within jets using either analytic resummation~\cite{Bain:2016clc} or the parton shower of a Monte Carlo event generator~\cite{Ernstrom:1996am}.
Quarkonium production in the parton shower, which can explain the lack of observed polarization~\cite{Baumgart:2014upa}, predicts that \jpsi mesons are rarely produced in isolation.
Consequently, it is of great interest to study the radiation produced in association with quarkonium states, {\em e.g.}\ \jpsi mesons in jets, to distinguish between these two different pictures of quarkonium production.

This Letter reports a study of \jpsi mesons produced in jets in the forward region of $pp$ collisions.
The fraction of the jet transverse momentum  carried by the \jpsi meson, $\z\equiv\pt(\jpsi)/\pt({\rm jet})$, is measured for \jpsi mesons produced promptly and for those produced in $b$-hadron decays.
The data sample corresponds to an integrated luminosity of 1.4\invfb collected at a center-of-mass energy of $\sqrt{s}=13\tev$ with the LHCb detector in 2016. Only events containing exactly one reconstructed $pp$ collision are used as these provide the best resolution on \ptj. The analysis is performed using jets clustered with the anti-$k_{\rm T}$ algorithm~\cite{antikt} using a distance parameter $R=0.5$ and within the following kinematic fiducial region:
jets are required to have $\ptj > 20\gev$ ($c=1$ throughout this Letter) in the pseudorapidity range $2.5 < \eta({\rm jet}) < 4.0$;
\jpsi mesons, which are reconstructed using the $\jpsi\to\mu^+\mu^-$ decay, must satisfy $2.0 < \eta(\jpsi) < 4.5$;
 and muons are required to have $\pt(\mu) > 0.5\gev$, $p(\mu) > 5\gev$, and $2.0 < \eta(\mu) < 4.5$.
No requirements are placed on the multiplicity of jets per event or particles per jet, so that jets consisting of only a \jpsi candidate are allowed.
This is the first measurement of \z in prompt \jpsi production at any experiment.

The \lhcb detector
is a single-arm forward
spectrometer covering the
range $2<\eta <5$, described in detail in Refs.~\cite{Alves:2008zz,LHCb-DP-2014-002}.
Simulated data samples are used to evaluate the muon reconstruction efficiency, the detector response for jet reconstruction, and to validate the analysis.
In the simulation, $pp$ collisions are generated using
\pythia~\cite{Sjostrand:2006za,*Sjostrand:2007gs}
 with a specific \lhcb
configuration~\cite{LHCb-PROC-2010-056}.  Decays of hadronic particles
are described by \evtgen~\cite{Lange:2001uf}, in which final-state
radiation is generated using \photos~\cite{Golonka:2005pn}. The
interaction of the generated particles with the detector, and its response,
are implemented using the \geant
toolkit~\cite{Allison:2006ve, *Agostinelli:2002hh} as described in
Ref.~\cite{LHCb-PROC-2011-006}.

The online event selection is performed by a trigger~\cite{LHCb-DP-2012-004}, which
consists of a hardware stage using information from the calorimeter and
muon systems,
followed by a software stage, which performs the \jpsi candidate reconstruction.
The hardware stage selects events with at least one dimuon candidate with $\sqrt{\pt(\mu^+)\pt(\mu^-)}$ greater than a threshold that varied between 1.3 and 1.5\gev during the 2016 data taking.
In the software stage, two muon candidates with $\pt(\mu) > 0.5\gev$ are required to form a \jpsi candidate whose invariant mass is within 150\mev of the known \jpsi mass\cite{PDG2016}.
Additional selection criteria are applied offline to the \jpsi candidates:
the tracks are required to satisfy stringent muon-identification criteria;
and the muon and \jpsi candidates are required to be within the fiducial region of this analysis, where the detector is well understood.

A new data-taking scheme was introduced by LHCb in 2015 that enables offline-like performance in the online system.
The alignment and calibration are performed in near real-time\cite{LHCb-PROC-2015-011}, and are available in the trigger reconstruction~\cite{Aaij:2016rxn}.
Furthermore, an increase in the online CPU resources makes it possible to run the offline track reconstruction in the online system.
This analysis is based on a data sample where all online-reconstructed particles in the event are stored, but most lower-level information is discarded, greatly reducing the event size.
This data-storage strategy makes it possible to record all events containing a \jpsi candidate without placing any requirements on $\pt(\jpsi)$, or on the displacement of the \jpsi decay from the primary vertex (PV).

Jet reconstruction is performed offline on this data sample by clustering the \jpsi candidates with charged and neutral particle-flow candidates~\cite{LHCb-PAPER-2013-058}, all reconstructed online, using the anti-$k_{\rm T}$ clustering algorithm as implemented in
\textsc{FastJet}~\cite{fastjet}.
This is the first LHCb analysis to use online-reconstructed particles that were not involved in the trigger decision.
The \jpsi candidates, rather than their component muons, are used in the clustering to prevent muons from a single \jpsi decay being clustered into separate jets.
Reconstructed jets with $\ptj > 15\gev$ and $2.5 < \eta({\rm jet}) < 4.0$ are kept for further analysis, where jets
in the \ptj range 15--20\gev
are retained for use in unfolding the detector response.
The $\eta({\rm jet})$ requirement, which is included in the fiducial region definition, ensures a nearly uniform resolution of 20--25\% on the \pt of the non-\jpsi component of the jet,
with minimal \pt dependence above 10\gev.
This is similar to the resolution achieved on data events~\cite{LHCb-PAPER-2013-058}
when using offline reconstruction
for \pt below 20\gev, but worse at higher \pt where the resolution in such events is about 15\%.
This degradation arises largely because calorimeter information not associated to particle-flow candidates is not stored in this data sample.

The jet momenta are not corrected for reconstruction bias.
Instead, the effect of the detector response on the \z distributions is removed using an unfolding procedure.
This involves first determining the reconstructed \jpsi yields in bins of $[\z,\ptj]$, then correcting them for detection efficiency.
Bin migration, which occurs largely due to the resolution on the non-\jpsi component of the jet, is accounted for by unfolding the $[\z,\ptj]$ distributions of corrected \jpsi yields using
an iterative Bayesian procedure~\cite{D'Agostini:1994zf,Adye:2011gm} (see the Supplemental Material to this Letter~\cite{Supp} for a detailed discussion of the unfolding).
Finally, the unfolded $[\z,\ptj]$ distributions are integrated for $\ptj > 20\gev$ to produce the measured \z spectra.
The binning scheme employs ten equal-width \z bins, and three \ptj bins of 15--20, 20--30, and $>30\gev$.

\begin{figure}
  \centering
  \includegraphics[width=0.49\textwidth]{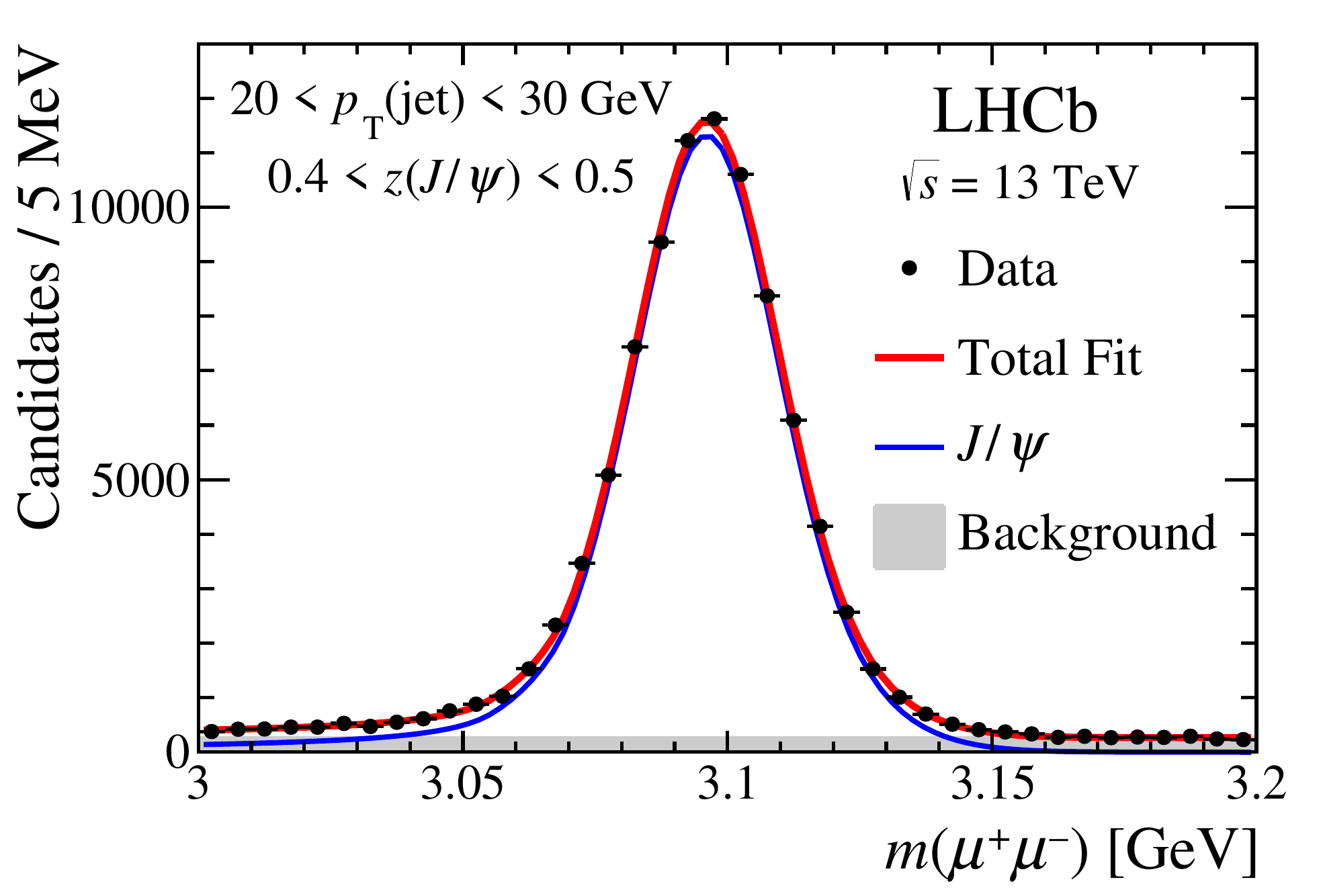}
  \caption{
    Example dimuon invariant-mass distribution with the fit result superimposed from the bin $[0.4 < \z < 0.5, 20 < \ptj < 30\gev]$.
    The signal is modeled as the sum of two Crystal Ball functions, while the background is described by an exponential function.
  }
  \label{fig:mfit}
\end{figure}

The yield of $\jpsi\to\mu^+\mu^-$ decays reconstructed in each $[\z,\ptj]$ bin, which includes \jpsi mesons produced promptly and in $b$-hadron decays, is determined from an unbinned maximum likelihood fit to the corresponding dimuon invariant-mass distribution.
The signal component is modeled as the sum of two Crystal Ball functions~\cite{Skwarnicki:1986xj} that share all shape parameters except the width.
The combinatorial background is described by an exponential function.
Both the signal and background shapes are allowed to vary in each bin independently.
An example of the invariant-mass distribution from one $[\z,\ptj]$ bin is shown in Fig.~\ref{fig:mfit} along with the fit result.
The total \jpsi signal yield in the data sample is almost two million.

\begin{figure}
  \centering
  \includegraphics[width=0.49\textwidth]{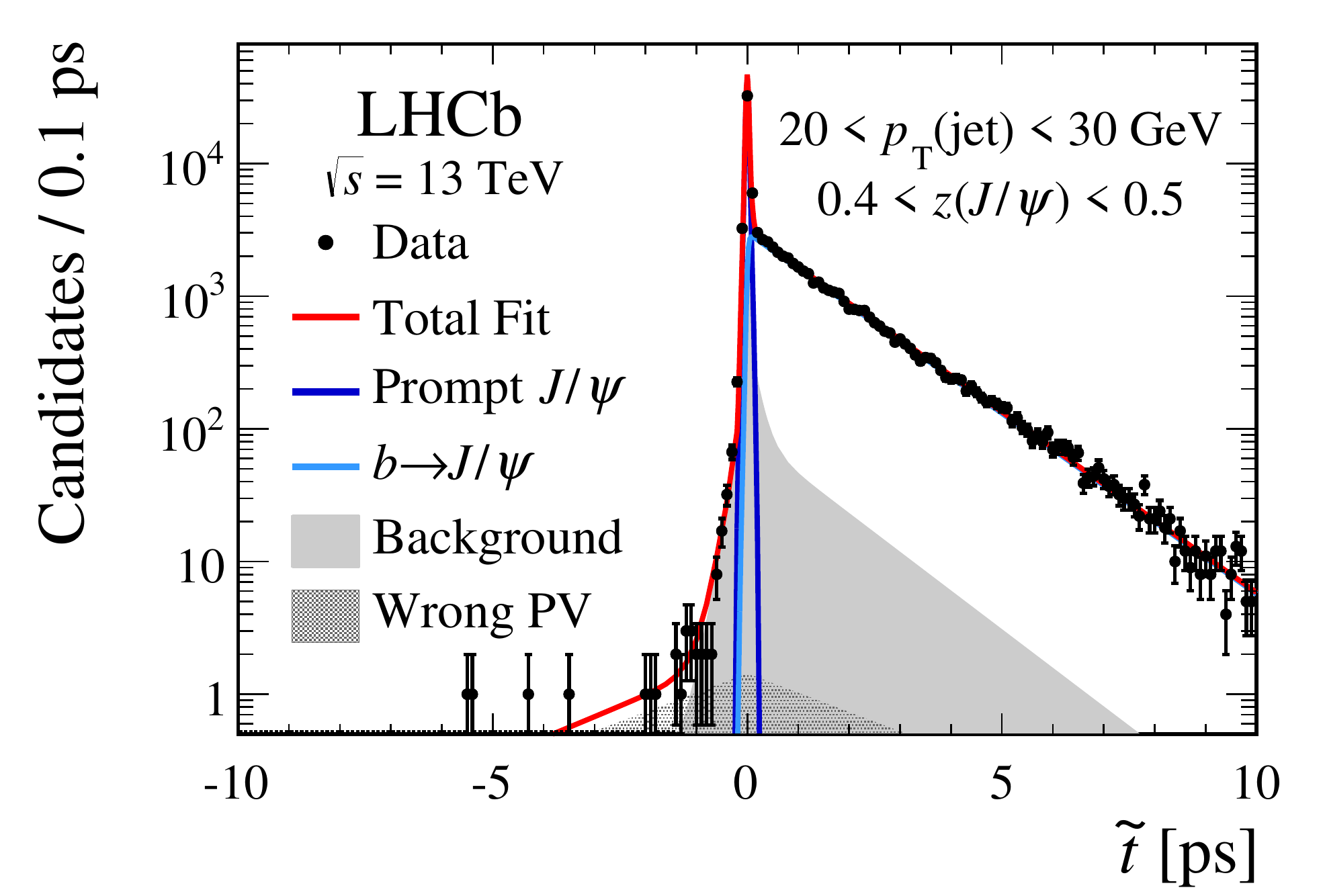}
  \includegraphics[width=0.49\textwidth]{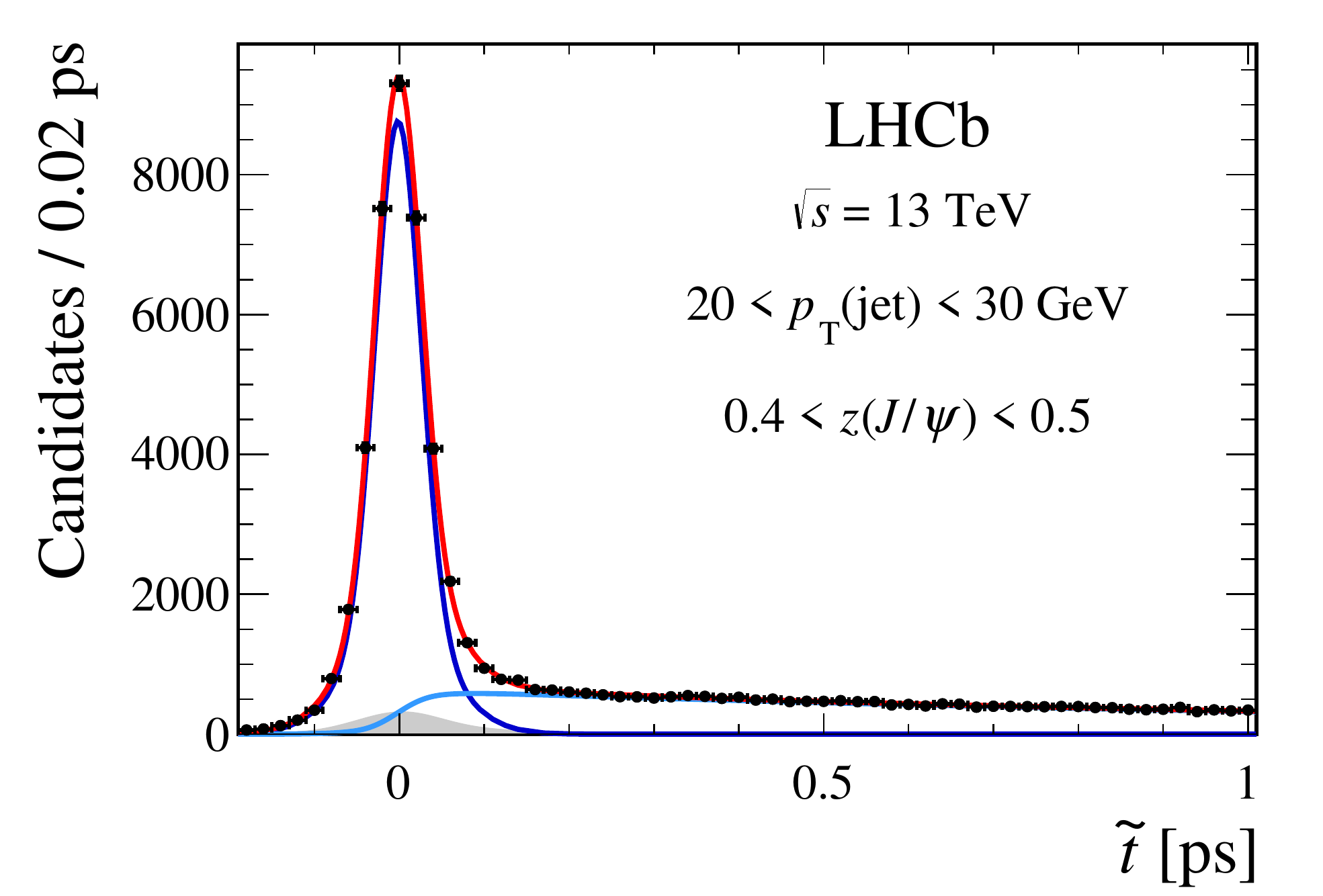}
  \caption{
    Example pseudo-decay-time distribution from the same bin as in Fig.~\ref{fig:mfit} with the fit result superimposed.
The right plot shows the $[-0.2,1]$\ps region on a linear scale.
  }
  \label{fig:tfit}
\end{figure}

The fraction of \jpsi mesons that originates from $b$-hadron decays is determined by fitting the distribution of the pseudo-decay-time
$\tz \equiv \lambda\, m(\jpsi) / p_{\rm L}(\jpsi)$,
where $\lambda$
denotes the difference in position along the beam axis between the \jpsi decay and primary vertices, $m(\jpsi)$ is the known \jpsi mass\cite{PDG2016}, and $p_{\rm L}(\jpsi)$ is the component of the \jpsi momentum longitudinal to the beam axis.
Only candidates with $|\tz| < 10\ps$, corresponding to about seven $b$-hadron lifetimes, and a mass consistent with the known \jpsi mass are used in these unbinned maximum likelihood fits.
The \tz distribution from one $[\z,\ptj]$ bin is shown in Fig.~\ref{fig:tfit}.
The prompt-\jpsi component is modeled by a Dirac $\delta$ function, while the $b$-hadron component is modeled by an exponential decay function with a variable lifetime parameter;
both are convolved with a double-Gaussian resolution function.
A long and nearly symmetric tail in the \tz distribution arises due to \jpsi candidates produced in additional $pp$ collisions that are not reconstructed. The shape of this component, the contribution of which is found to be $\mathcal{O}(0.1\%)$ in all bins, is modeled by constructing the distribution with \tz  calculated using \jpsi and PV candidates from different data events.
Finally, the shape of the non-\jpsi component in  each %$[\z,\ptj]$
bin is parametrized using an empirical function obtained from a fit to the \tz distribution observed in the $m(\mu^+\mu^-)$ sidebands, while its normalization is fixed from  the $m(\mu^+\mu^-)$ fit in the bin.
 The fraction of \jpsi mesons that are produced in $b$-hadron decays is determined to be in the range 20--60\%, depending on $[\z,\ptj]$ bin.

The \jpsi yields are corrected for detection efficiency by applying per-candidate weights of $\varepsilon_{{\rm tot}}^{-1}$, where $\varepsilon_{{\rm tot}}$ is the total detection efficiency determined as the product of the reconstruction, selection, and trigger efficiencies.
The use of per-candidate weights within a fiducial region where the efficiency is nonzero throughout produces accurate efficiency-corrected yields without requiring knowledge of the $\jpsi\to\mu^+\mu^-$ angular distribution or, equivalently, the \jpsi polarization.
The weights, which are similar for nearly all candidates, are rarely greater than 5 and never greater than 20.
Consequently, there is negligible impact on the statistical variance due to the use of weighted candidates, since the vast majority of events in each $[\z,\ptj]$ bin contribute nearly equally.

The muon reconstruction efficiency is obtained from simulation in bins of $[p(\mu),\eta(\mu)]$.
Scale factors that correct for discrepancies between the data and simulation are determined using a data-driven tag-and-probe approach on an independent sample of $\jpsi\to\mu^+\mu^-$ decays~\cite{LHCb-DP-2013-002}.
A small $\pt(\jpsi)$-dependent correction is applied to the yields of \jpsi mesons produced in $b$-hadron decays to account for a drop in the efficiency at large $b$-hadron flight distances.
Within the fiducial region of this analysis, the \jpsi reconstruction efficiency is on average about 90\%.

The dominant contribution to the selection inefficiency is from the muon-identification performance, which is measured in bins of $[\pt(\mu),\eta(\mu)]$ using a highly pure calibration data sample of $\jpsi\to\mu^+\mu^-$ decays.
The efficiency of selecting a reconstructed \jpsi candidate varies from 80\% for $\z \lesssim 0.1$ to nearly 100\% for $\z \gtrsim 0.5$.
 The trigger efficiency is measured in bins of $[\sqrt{\pt(\mu^+)\pt(\mu^-)},\eta(\jpsi)]$ using a subset of this \jpsi calibration sample.
Events selected by the hardware trigger independently of the \jpsi candidate, \eg due to the presence of a high-\pt hadron, are used to determine the trigger efficiency directly from the data.
The fraction of \jpsi candidates in each $[\sqrt{\pt(\mu^+)\pt(\mu^-)},\eta(\jpsi)]$ bin that are selected by the dimuon hardware trigger gives the efficiency, which is about 40\% on average for $\z \lesssim 0.1$ and 80\% for $\z \gtrsim 0.5$.

The effects of $[\z,\ptj]$ bin migration, which are predominantly due to the detector response to the non-\jpsi component of the jet,
are corrected for using an unfolding technique~\cite{D'Agostini:1994zf,Adye:2011gm,Supp}.
The detector-response matrices for \jpsi mesons produced promptly and in $b$-hadron decays are dissimilar for two reasons:
the \pt-dependent particle multiplicities are different, and the undetected momentum carried by \Kz and \Lz particles is, on average, larger for jets that contain a $b$-hadron decay.
The \ptj-dependent mean and width of the reconstructed particle multiplicity distributions for jets in simulation are adjusted to match those observed in data.
The detector response is studied using the \pt-balance distribution of $\ptj/\pt(Z)$ in nearly back-to-back $Z+$jet events using the same data-driven technique as in Ref.~\cite{LHCb-PAPER-2013-058}.
Small adjustments are applied to the \ptj scale and resolution in simulation to obtain the best agreement with data.
The unfolding matrix for jets that contain a prompt \jpsi meson is shown in Fig.~\ref{fig:unfold_matrix}, while the corresponding matrix for $b$-hadron production
is provided in the Supplemental Material~\cite{Supp}.

\begin{figure}
  \centering
  \includegraphics[width=0.49\textwidth]{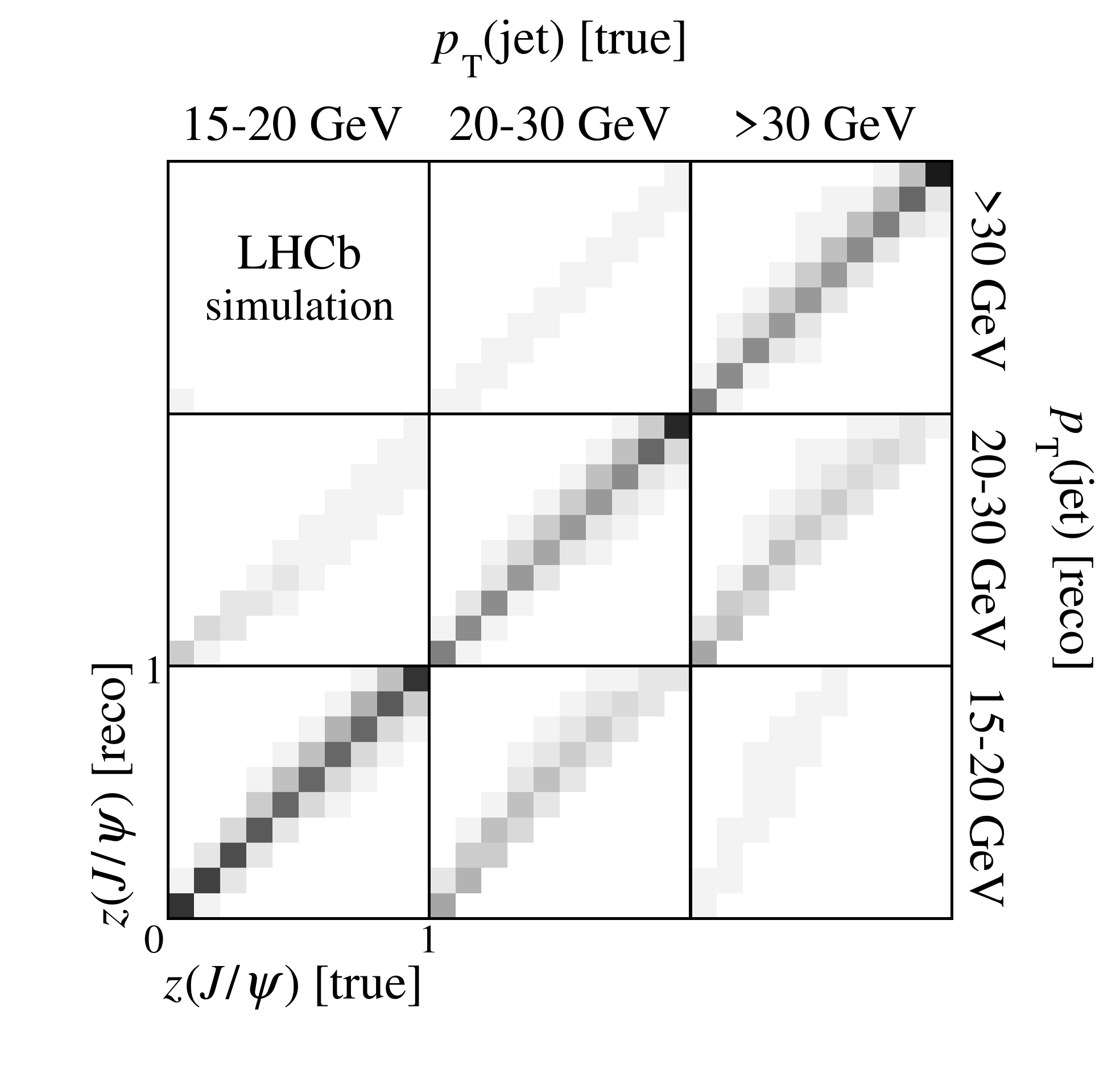}
  \caption{
   The four-dimensional detector-response matrix for prompt \jpsi production. The shading represents the bin-to-bin migration probabilities ranging from (white) 0 to (black) 1, with the lightest shade of gray corresponding to a probability of 0.1--0.3\%.
Jets whose true \ptj is above 20\gev but whose reconstructed \ptj is below 15\gev, or {\em vice versa}, are included in the unfolding but not shown graphically.
  }
  \label{fig:unfold_matrix}
\end{figure}

Systematic uncertainties on the \z distributions apply to both the prompt and $b$-hadron production modes.
Uncertainty on the \jpsi yields arises from the efficiency corrections and from possible mismodeling of the components in the invariant-mass and pseudo-decay-time fits.
The uncertainty on each component of the total efficiency is assessed by
repeating the data-driven efficiency studies
on simulated events, where
the difference between the true and efficiency-corrected \jpsi yields in bins of $[\pt(\jpsi),\eta(\jpsi)]$ is used to determine the systematic uncertainty.
The relative uncertainty on the reconstruction efficiency is determined to be 2\%, which includes the unknown \jpsi polarization.
The relative uncertainties on the trigger and selection efficiencies are in the ranges 2--5\% and 0--2\%, respectively, depending on $[\z,\ptj]$ bin.

The uncertainty on the total \jpsi yield obtained from the invariant-mass fits (1\%) is studied by replacing the nominal signal and background models with single Crystal Ball and quadratic functions, respectively.
The relative uncertainty on the fraction of \jpsi mesons produced in $b$-hadron decays (1\%) is determined by comparing the fit results obtained from simulated \tz distributions to the true fractions.
Potential mismodeling of the non-\jpsi and wrong-PV components is found to contribute negligible uncertainty.
The total relative uncertainty on the \jpsi yields is 3--6\% depending on $[\z,\ptj]$ bin,
which corresponds to a bin-dependent absolute uncertainty on \z
of 0.001--0.005.

The uncertainty associated with the detector response to the non-\jpsi component of the jet is studied by building alternative unfolding matrices, where the \pt scale and resolution are varied within the uncertainties obtained from the data-driven \pt-balance study of $Z+$jet events.
The data are unfolded using these alternative matrices, with the differences in the \z distribution used to assign \z-dependent absolute uncertainties of 0.001--0.014.
The \ptj and \z spectra used to generate the unfolding matrices, along with the unfolding procedure itself, are also potential sources of uncertainty.
These are studied by simulating data samples similar to the experimental data, then unfolding them using response matrices constructed from \ptj and \z distributions that are different from those used to generate the samples.
Based on these studies, an uncertainty of 0.01 is assigned to each \z bin due to unfolding.
Finally, the uncertainties due to the fragmentation model and due to the \Kz and \Lz components of the jet are found to be negligible.
The total absolute systematic uncertainty in each \z bin, which dominates over the statistical one, is 0.010--0.015.

\begin{figure}
  \centering
  \includegraphics[width=0.49\textwidth]{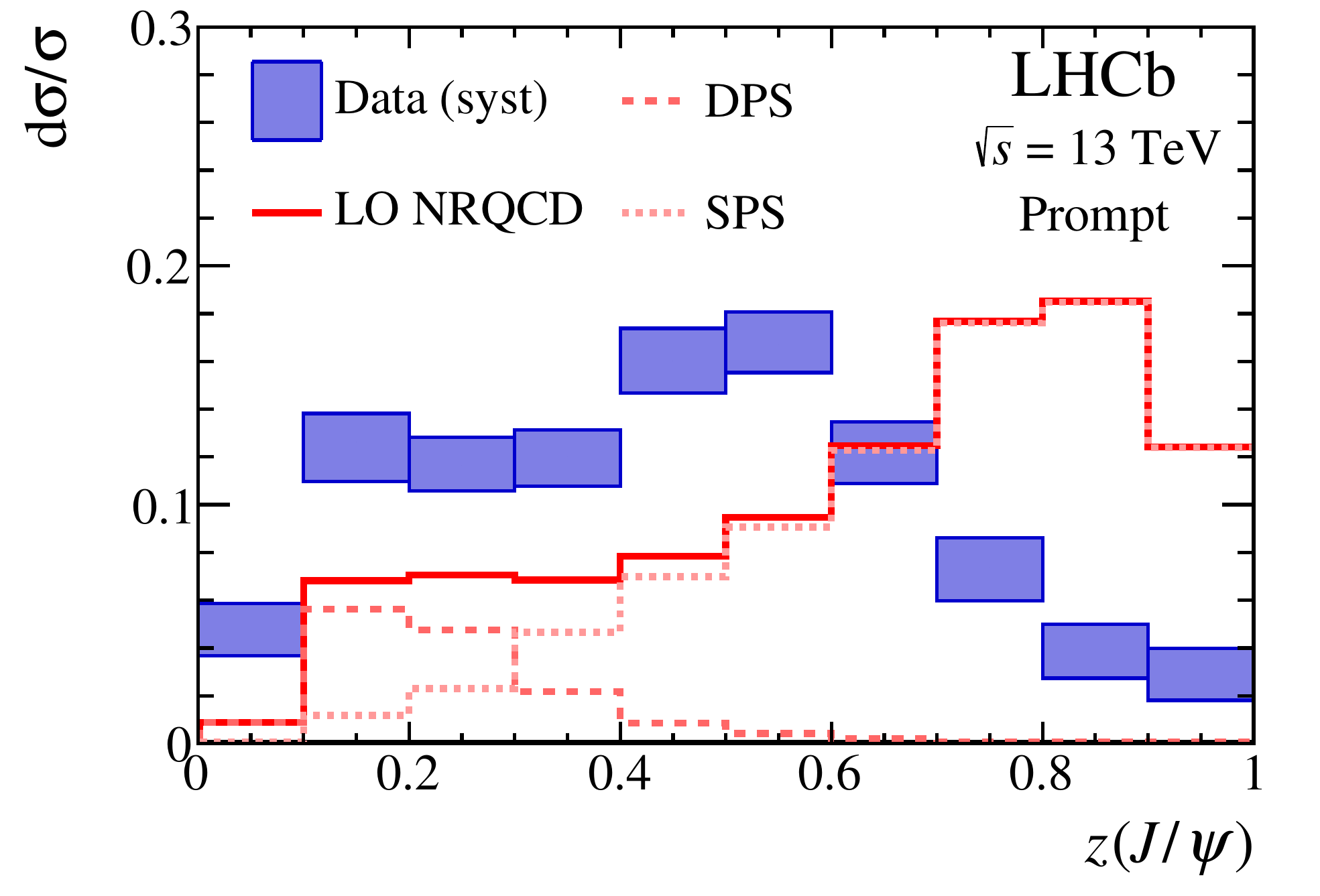}
  \includegraphics[width=0.49\textwidth]{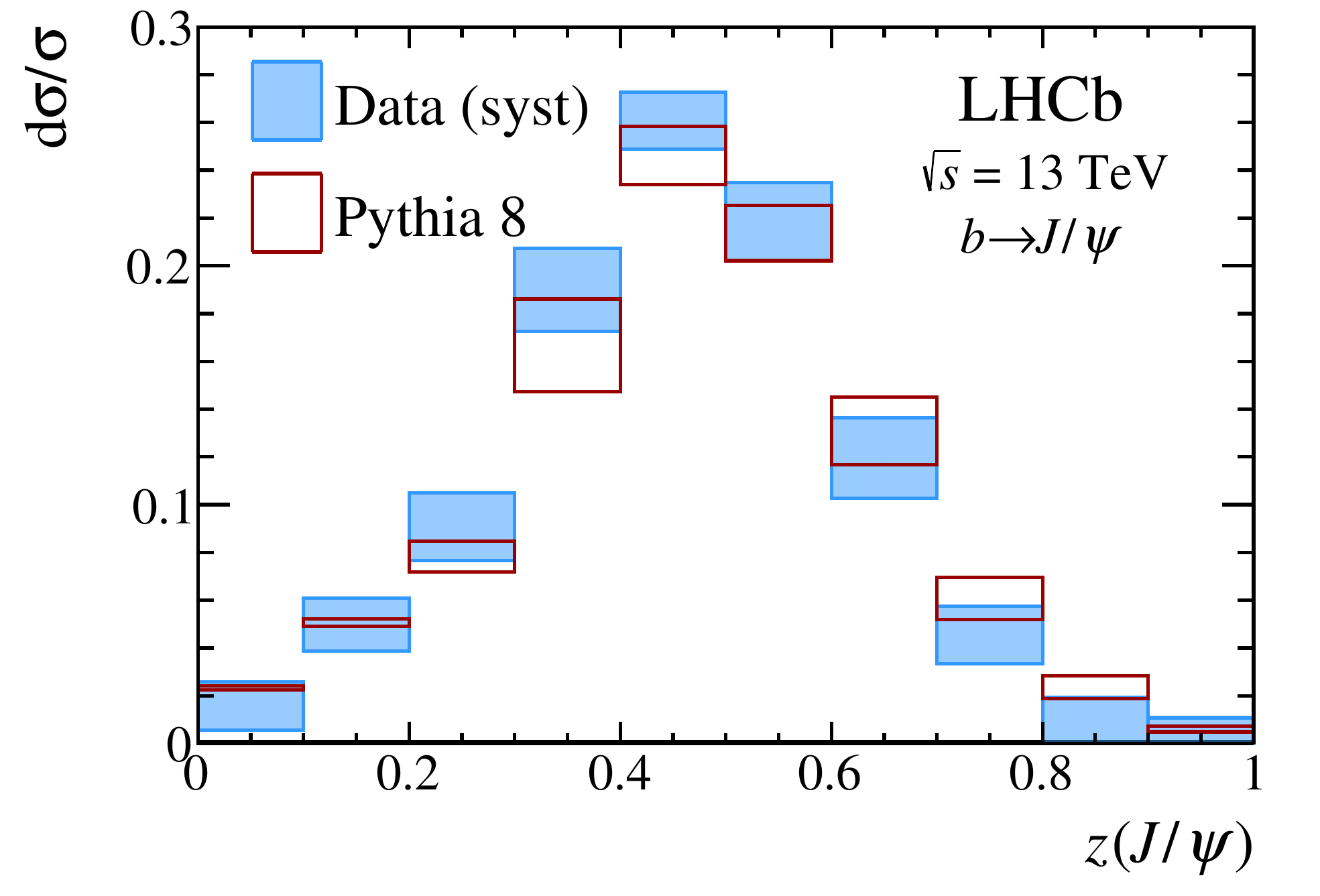}
  \caption{
    Measured normalized \z distributions for \jpsi mesons produced (left) promptly and (right) in $b$-hadron decays, compared to predictions obtained from \pythia.
    The statistical uncertainties are negligible.
    The (DPS) double and (SPS) single parton scattering contributions to the prompt prediction are also shown (the DPS effective cross section in \pythia is 31\,mb).
}
  \label{fig:z}
\end{figure}

The measured normalized \z distributions for \jpsi mesons produced promptly and for those produced in $b$-hadron decays are shown in Fig.~\ref{fig:z} (the numerical values are provided in Ref.~\cite{Supp}).
The $b$-hadron results are consistent with the \pythia prediction~\cite{Supp,Monash}, where the uncertainty shown is due to $b$-quark fragmentation~\cite{Nason:1999ta,Khachatryan:2016wqo} (other sources of uncertainty are ignored~\cite{Supp}).
The prompt-\jpsi results do not agree with the leading-order (LO) NRQCD-based prediction as implemented in \pythia,
which includes both color-octet and color-singlet mechanisms using long-distance matrix elements determined empirically~\cite{Supp}.
At small \z, \pythia predicts that most of \ptj arises from a parton-parton scatter other than the one that produced the \jpsi meson.
The dominant source of uncertainty on the prompt-\jpsi prediction at large \z is due to the underlying event;
however, since no rigorous method exists for determining this uncertainty,
no uncertainty is assigned to the prediction.
Given that the underlying event at LHCb is well described by \pythia, {\em e.g.},\ the energy flow is accurately predicted at the 5\% level~\cite{LHCb-PAPER-2012-034}, the prompt-\jpsi results
%in Fig.~\ref{fig:z}
cannot be reconciled with this prediction.
Furthermore, LO and partial next-to-leading-order (NLO*) calculations in both the color-singlet and color-octet models similarly fail to describe the data~\cite{Supp,Shao:2015vga}.

Prompt \jpsi mesons in data are observed to be much less isolated than predicted, which qualitatively agrees with the alternative picture of quarkonium production presented in Ref.~\cite{Bain:2016clc} (after this Letter was submitted, Ref.~\cite{Bain:2017wvk} demonstrated quantitative agreement).
The lack of isolation observed for prompt \jpsi production may be related to the long-standing quarkonium polarization puzzle.
If high-\pt\,\!\jpsi mesons are predominantly produced within parton showers, rather than directly in parton-parton scattering, then the observed lack of both polarization and isolation could be explained~\cite{Baumgart:2014upa}.
Future related measurements of \jpsi production in jets should help shed light on the nature of quarkonium production~\cite{Ilten:2017rbd,Kang:2017yde}.

In summary, the production of \jpsi mesons in jets is studied using $pp$-collision data collected by LHCb at $\sqrt{s} = 13\tev$ in the fiducial region: $\ptj > 20\gev$ and $2.5 < \eta({\rm jet}) < 4.0$; $2.0 < \eta(\jpsi) < 4.5$; and $\pt(\mu) > 0.5\gev$, $p(\mu) > 5\gev$, and $2.0 < \eta(\mu) < 4.5$.
The fraction of the jet \pt carried by the \jpsi meson is measured for \jpsi mesons produced promptly and for those produced in $b$-hadron decays.
 The observed distribution for \jpsi mesons produced in $b$-hadron decays is consistent with the \pythia prediction; however,
the prompt-\jpsi results do not agree with predictions based on fixed-order NRQCD as implemented in \pythia.

\section*{Acknowledgments}

\noindent
We express our gratitude to our colleagues in the CERN
accelerator departments for the excellent performance of the LHC. We
thank the technical and administrative staff at the LHCb
institutes. We acknowledge support from CERN and from the national
agencies: CAPES, CNPq, FAPERJ and FINEP (Brazil); NSFC (China);
CNRS/IN2P3 (France); BMBF, DFG and MPG (Germany); INFN (Italy);
FOM and NWO (The Netherlands); MNiSW and NCN (Poland); MEN/IFA (Romania);
MinES and FASO (Russia); MinECo (Spain); SNSF and SER (Switzerland);
NASU (Ukraine); STFC (United Kingdom); NSF (USA).
We acknowledge the computing resources that are provided by CERN, IN2P3 (France), KIT and DESY (Germany), INFN (Italy), SURF (The Netherlands), PIC (Spain), GridPP (United Kingdom), RRCKI and Yandex LLC (Russia), CSCS (Switzerland), IFIN-HH (Romania), CBPF (Brazil), PL-GRID (Poland) and OSC (USA). We are indebted to the communities behind the multiple open
source software packages on which we depend.
Individual groups or members have received support from AvH Foundation (Germany),
EPLANET, Marie Sk\l{}odowska-Curie Actions and ERC (European Union),
Conseil G\'{e}n\'{e}ral de Haute-Savoie, Labex ENIGMASS and OCEVU,
R\'{e}gion Auvergne (France), RFBR and Yandex LLC (Russia), GVA, XuntaGal and GENCAT (Spain), Herchel Smith Fund, The Royal Society, Royal Commission for the Exhibition of 1851 and the Leverhulme Trust (United Kingdom).

\setboolean{inbibliography}{true}
%\bibliographystyle{LHCb}
%\mciteErrorOnUnknownfalse
\bibliography{paper.bbl}

\newpage

\section*{Supplemental Material}

\subsection*{Unfolding}

\begin{figure}[h!]
  \centering
  \includegraphics[width=0.6\textwidth]{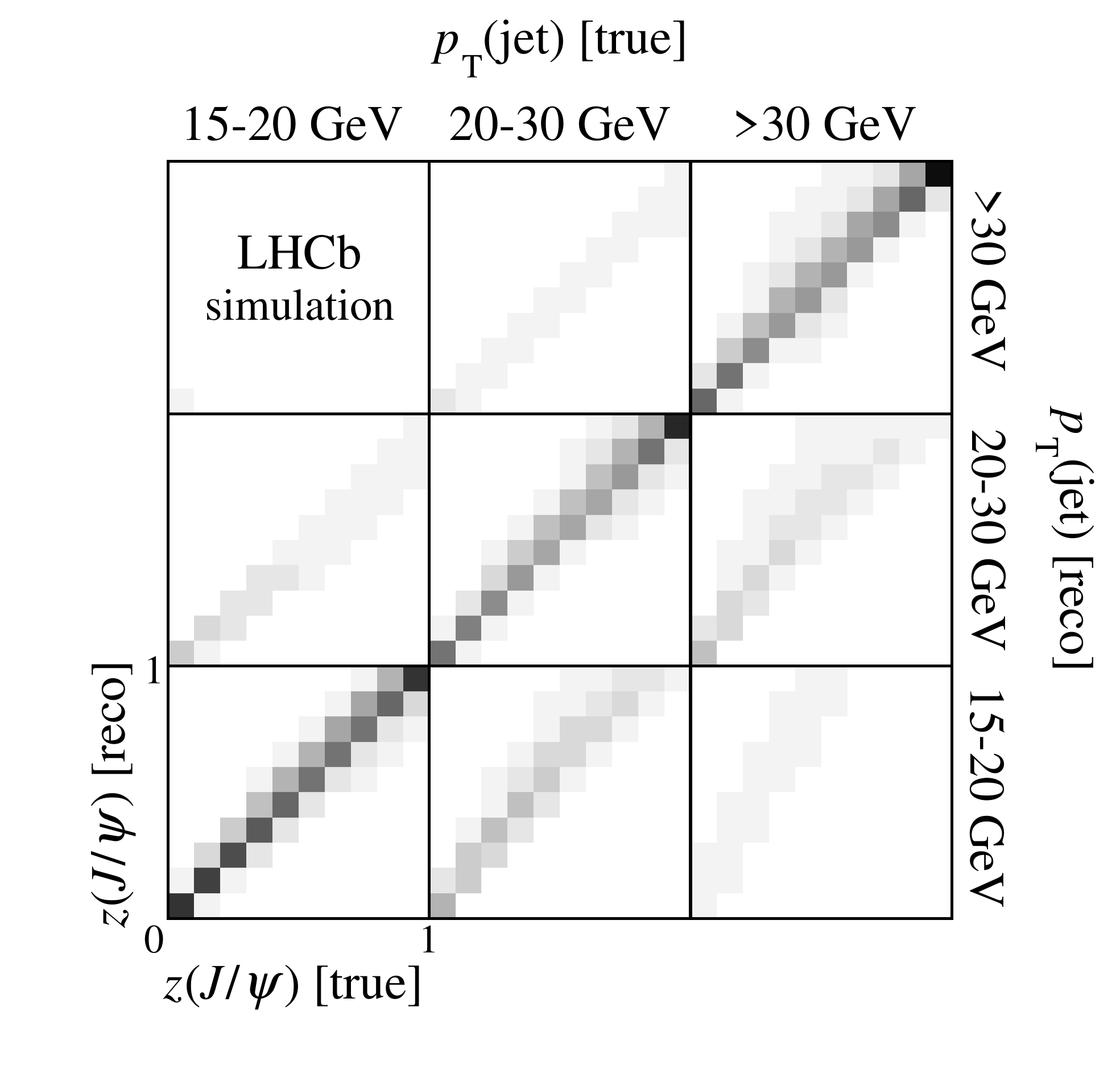}
  \caption{
    Analog to Fig.~3 in the Letter for \jpsi mesons produced in $b$-hadron decays.
  }
  \label{fig:unfold_matrices}
\end{figure}

The unfolding is performed using an iterative Bayesian approach~\cite{D'Agostini:1994zf} as implemented in \textsc{RooUnfold}~\cite{Adye:2011gm}.
The unfolding matrices for both the prompt and $b$-hadron-decay cases are sensitive to the \z distributions used to construct them.
In the initial unfolding step, both matrices are generated under the assumption of a uniform true \z distribution.
Three iterations of the unfolding technique are performed using these matrices, producing initial unfolded  \z distributions.
Next, these unfolded \z distributions are used to construct updated unfolding matrices, which are then used to perform three further iterations of the unfolding technique on the data, producing improved \z distributions.
This process is repeated one more time to obtain the final unfolded \z distributions presented in the Letter; therefore, in total, 9 iterations of the unfolding technique are performed using 3 unfolding matrices for both prompt  and $b$-hadron production.
The procedure is terminated after three super-iterations for two reasons: it is found to converge, {\em i.e.}\ the differences between the input and output \z distributions are $\mathcal{O}(0.001)$; and in studies performed on simulated data samples, no gain in accuracy is observed using additional iterations.
The unfolding matrices for both prompt and $b$-hadron production used in the final super-iterations are shown in Fig.~3 in the Letter and in Fig.~\ref{fig:unfold_matrices}, respectively.

\clearpage

\subsection*{Theory Predictions}

\newcommand{\helaconia}{\textsc{HelacOnia}\xspace}
\newcommand{\pth}{\ensuremath{\hat{p}_\mathrm{T}}}

All \pythia predictions are generated using {\sc Pythia}~8.212 and {\sc EvtGen}~1.04.
Radiative \jpsi decays are handled by \photos using the {\tt PHOTOS VLL} setting, while the standard {\sc EvtGen} decay tables are used for all other decays.
The production of prompt \jpsi mesons is simulated in \pythia with the flag {\tt Charmonimum:all=on}, while the production of \jpsi mesons from  $b$-hadron decays is done with {\tt HardQCD:all=on}.

Prompt \jpsi production in NRQCD is calculated using a summed expansion of Fock states. The leading term is color-singlet \jpsi production, while the sub-leading terms describe color-octet $c\bar{c}$ production in association with a gluon. Each term is factorized into a perturbatively calculated short-distance matrix element (SDME) and an empirically determined long-distance matrix element (LDME). Within \pythia, both the leading color-singlet ${}^3S_1^{(1)}$ and subleading color-octet ${}^3S_1^{(8)}$, ${}^1S_0^{(8)}$, and ${}^3P_J^{(8)}$ states are used to calculate prompt \jpsi production. The unpolarized SDMEs for each state are calculated at tree level, while the LDMEs are taken from Ref.~\cite{Nason:1999ta}. Since the \pythia SDMEs are LO, the corresponding LDMEs are taken from LO fits. Feed-down from higher mass charmonium states is included, also calculated with both color-singlet and color-octet contributions.

The \pythia parton shower is applied to color-octet states, but not color-singlet states, where a quark-splitting kernel with twice the standard probability is used. After showering, all color-octet states are forced to decay isotropically into a color-singlet state with an associated soft gluon. The mass splitting for this decay is $200\mev$. Consequently, color-singlet production in \pythia is isolated, while color-octet production is accompanied by soft radiation which has only a small effect on \z. A larger effect comes from multi-parton interactions coincident with the \jpsi, shifting \z to lower values.
%Since the \jpsi cross-section diverges as $\pth \to 0$, regularized via the standard $\pt$ damping in \pythia, the contribution to low \z from DPS with a high \pt jet is non-negligible.

In Fig.~4 of the Letter, the full LO NRQCD prediction is calculated using all $pp$ collisions where a charmonium ($c\bar{c}$) system is produced with $\pth(c\bar{c}) > 2\gev$.
The DPS contribution is calculated using $pp$ collisions where both a $c\bar{c}$ and a dijet ($jj$) system are produced in separate parton-parton scatters, with $\hat{p}_T(c\bar{c}) > 2\gev$ and $\pth(jj) > 10\gev$. Additionally, $\Delta R < 0.5$ is required between the final-state \jpsi and a parton from the $jj$ system.
The SPS prediction is defined as the difference between the full and DPS calculations.
The shape and normalization of the SPS contribution is validated against a full LO NRQCD prediction with $\pth(c\bar{c}) > 10\gev$.
Multi-parton interactions are included for all \pythia calculations.

\begin{figure}
  \centering
  \includegraphics[width=0.49\textwidth]{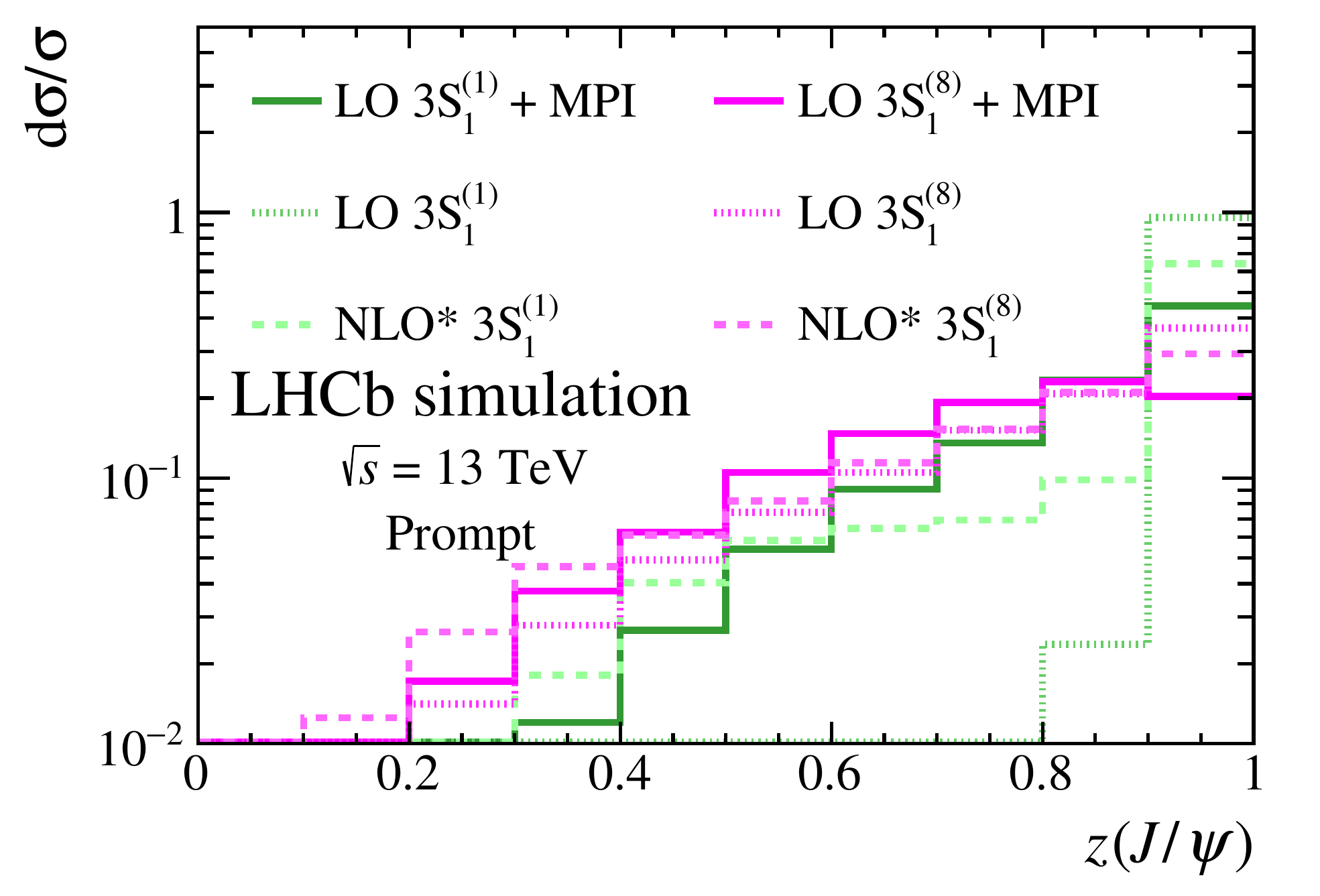}
    \includegraphics[width=0.49\textwidth]{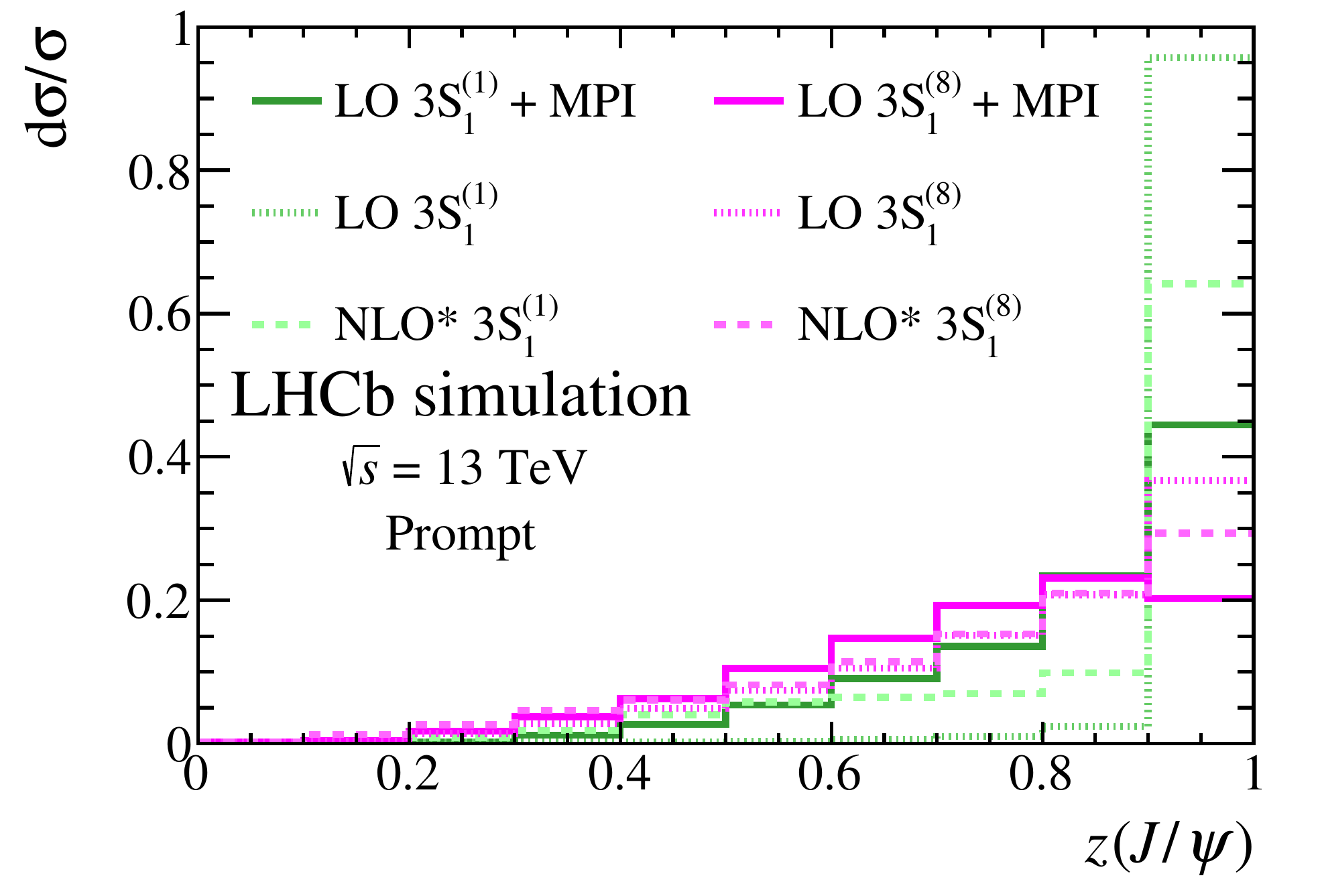}\\
    \includegraphics[width=0.49\textwidth]{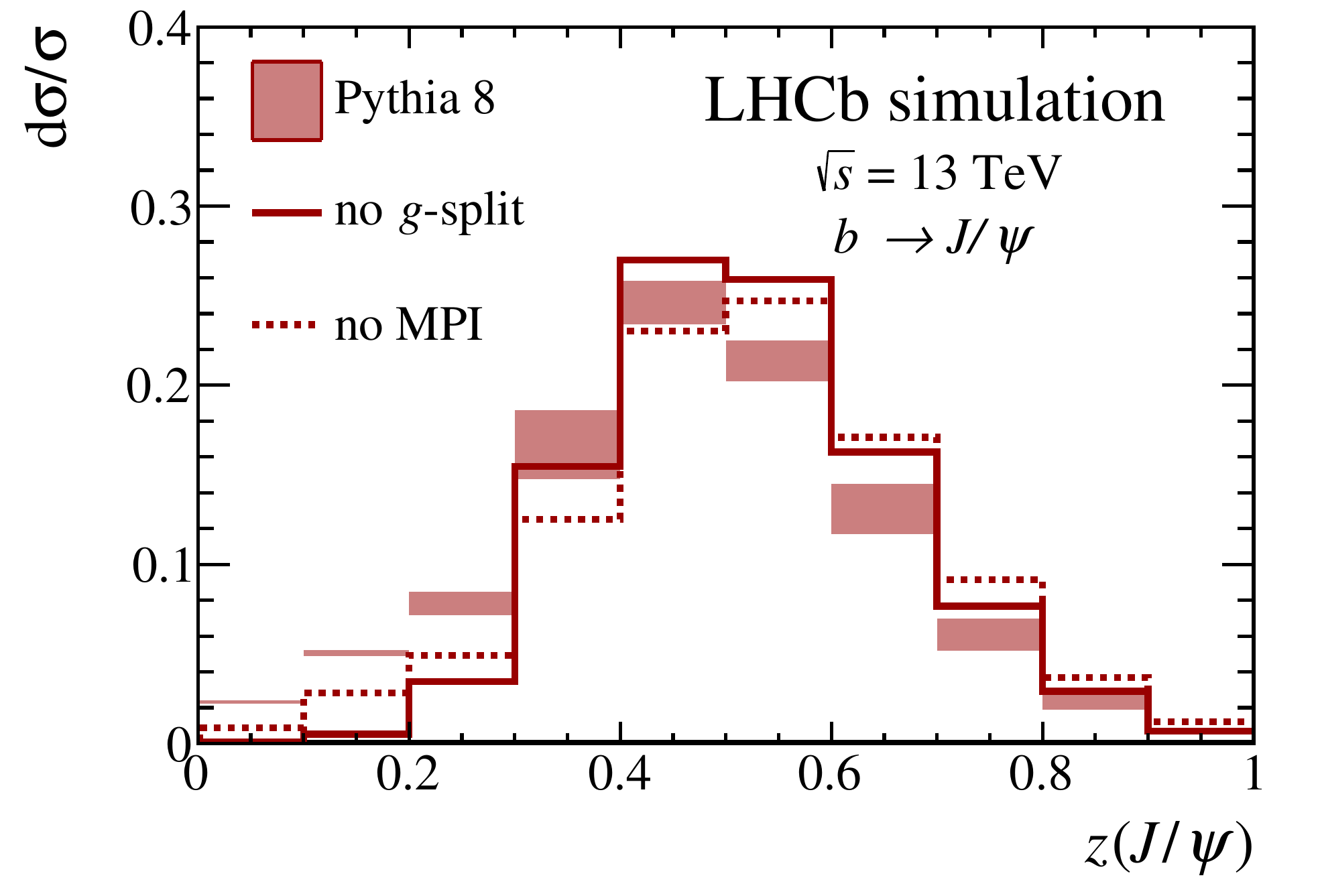}
  \caption{
  (top) Prompt-\jpsi calculations of \z for color-singlet and color-octet production, with and without multi-parton interactions (MPI), {\em i.e.}\ the soft underlying event (the left plot is a log-scale version of the right).
    (bottom) Nominal \z prediction for \jpsi mesons produced in $b$-hadron decays, compared to predictions (solid line) without gluon splitting and (dashed line) without MPI.}
  \label{fig:ztheory}
\end{figure}

Further SPS calculations have been performed using \helaconia~\cite{Shao:2015vga} for color-singlet $3\mathrm{S}_1^{(1)}$ and color-octet $3\mathrm{S}_1^{(8)}$ \jpsi production at LO and NLO*, which includes real but not virtual corrections. The \helaconia calculations are performed with $\pth(c\bar{c}) > 10\gev$ and interfaced with the \pythia parton shower.
Because of technical limitations, multi-parton interactions are not included. The LO \helaconia predictions are consistent with equivalent \pythia predictions.
In Fig.~\ref{fig:ztheory} the LO and NLO* \helaconia predictions are compared to the LO \pythia predictions with multi-parton interactions.
%Full NLO* NRQCD predictions, including the remaining color-octet states and feed-down, are not provided as the qualitative behaviour of NLO* with respect to LO is clear.

Production of \jpsi mesons from  $b$-hadron decays primarily depends on $b$-quark fragmentation. A systematic analysis of the fragmentation in \textsc{Pythia}, specifically of the Bowler parameter $r_b$, was performed in the $\mathrm{Z}2^*$ tune by CMS~\cite{Khachatryan:2016wqo} using \textsc{Pythia 6}. The $r_b$ parameter was found to be to $0.591 ^{+0.216}_{-0.274}$, where the uncertainty corresponds to one standard deviation.  In the present analysis, \pythia with the Monash tune~\cite{Monash} is used for theory predictions and so the $r_b$ uncertainty from the $\mathrm{Z2}^*$ tune is assigned to the Monash $r_b$ value of 0.855. The error band on the nominal \pythia prediction shown in Fig.~4 in the Letter is determined from this $r_b$ uncertainty.

The \z distribution for \jpsi mesons from $b$-hadron decays is also sensitive to the amount of $g \to b\bar{b}$ splitting and the soft underlying event. The nominal \pythia prediction includes both gluon splitting and the soft underlying event. Figure~\ref{fig:ztheory} compares the nominal prediction to a prediction without gluon splitting ({\tt HardQCD:hardbbbar=on}) and a prediction without the underlying event ({\tt PartonLevel:MPI=off}). Neglecting either of these effects can result in a \z spectrum considerably harder than the nominal prediction.
Although the uncertainties on these effects are neglected in this analysis, their sizes suggest that the associated uncertainties are not small;  thus, the overall theory error on \z is probably underestimated.
However, this would not change the conclusion: data and theory are consistent for \jpsi mesons produced in $b$-hadron decays.

\clearpage

\subsection*{Numerical Results}

Numerical results are provided in Tables~\ref{tab:results}--\ref{tab:bcov}.
The correlation matrices are based on the systematic uncertainties (the statistical uncertainties are negligible).
Systematic uncertainties that are not correlated between the ten \z bins obtain a correlation coefficient of $-1/9$ from the normalization condition.
Bin-to-bin correlations that arise from the unfolding, which are predominantly due to the \ptj scale uncertainty, are evaluated in the same studies used to assess these uncertainties; {\em i.e.}\ the  correlated shifts observed when unfolding the data using alternative unfolding matrices are used to assign the correlation coefficients.

\begin{table}[h!]
  \begin{center}
    \caption{\label{tab:results} Summary of the measured ${\rm d}\sigma/\sigma$ results in bins of \z, where uncertainties are systematic (statistical uncertainties are negligible).
    }
      \begin{tabular}{ccc}
        \toprule
        \z & prompt \jpsi & $b\to\jpsi$ \\
        \midrule
0.0--0.1 & $0.047\pm0.011$ & $0.016\pm0.010$ \\
0.1--0.2 & $0.126\pm0.014$ & $0.050\pm0.011$ \\
0.2--0.3 & $0.116\pm0.011$ & $0.090\pm0.014$ \\
0.3--0.4 & $0.120\pm0.012$ & $0.190\pm0.017$ \\
0.4--0.5 & $0.160\pm0.014$ & $0.261\pm0.012$ \\
0.5--0.6 & $0.167\pm0.013$ & $0.219\pm0.016$ \\
0.6--0.7 & $0.122\pm0.013$ & $0.120\pm0.017$ \\
0.7--0.8 & $0.074\pm0.013$ & $0.045\pm0.012$ \\
0.8--0.9 & $0.039\pm0.011$ & $0.010\pm0.010$ \\
0.9--1.0 & $0.029\pm0.011$ & $0.001^{+0.010}_{-0.001}$ \\
 \bottomrule
      \end{tabular}
  \end{center}
\end{table}

%\clearpage

\begin{table}[h!]
%  \begin{center}
    \caption{\label{tab:pcov} Correlation matrix for prompt \jpsi production.}
    \hspace{-0.75in}
      \begin{tabular}{c|cccccccccc}
        \toprule
        \z & 0.0--0.1 & 0.1--0.2 & 0.2--0.3 & 0.3--0.4 & 0.4--0.5 & 0.5--0.6 & 0.6--0.7 & 0.7--0.8 & 0.8--0.9 & 0.9--1.0 \\
        \midrule
0.0--0.1 & \phantom{$-$}1.00 & \phantom{$-$}$0.03$  & $-0.03$  & $-0.02$  & \phantom{$-$}$0.00$  & $-0.18$  & $-0.17$  & $-0.16$  & $-0.19$  & $-0.16$ \\
0.1--0.2 & &  \phantom{$-$}1.00 & \phantom{$-$}$0.02$  & \phantom{$-$}$0.09$  & \phantom{$-$}$0.24$  & $-0.32$  & $-0.36$  & $-0.30$  & $-0.22$  & $-0.17$ \\
0.2--0.3 & &  &  \phantom{$-$}1.00 & $-0.04$  & $-0.01$  & $-0.16$  & $-0.17$  & $-0.15$  & $-0.16$  & $-0.14$ \\
0.3--0.4 & &  &  &  \phantom{$-$}1.00 & \phantom{$-$}$0.08$  & $-0.27$  & $-0.25$  & $-0.26$  & $-0.22$  & $-0.20$ \\
0.4--0.5 & &  &  &  &  \phantom{$-$}1.00 & $-0.33$  & $-0.40$  & $-0.38$  & $-0.21$  & $-0.17$ \\
0.5--0.6 & &  &  &  &  &  \phantom{$-$}1.00 & \phantom{$-$}$0.16$  & \phantom{$-$}$0.12$  & \phantom{$-$}$0.05$  & \phantom{$-$}$0.02$ \\
0.6--0.7 & &  &  &  &  &  &  \phantom{$-$}1.00 & \phantom{$-$}$0.21$  & \phantom{$-$}$0.05$  & \phantom{$-$}$0.00$ \\
0.7--0.8 & &  &  &  &  &  &  &  \phantom{$-$}1.00 & \phantom{$-$}$0.05$  & \phantom{$-$}$0.01$ \\
0.8--0.9 & &  &  &  &  &  &  &  &  \phantom{$-$}1.00 & \phantom{$-$}$0.01$ \\
0.9--1.0 & &  &  &  &  &  &  &  &  &  \phantom{$-$}1.00\\
        \bottomrule
      \end{tabular}
 % \end{center}
\end{table}

\begin{table}[h!]
%  \begin{center}
    \caption{\label{tab:bcov} Correlation matrix for \jpsi production in $b$-hadron decays.}
    \hspace{-0.75in}
      \begin{tabular}{c|cccccccccc}
        \toprule
        \z & 0.0--0.1 & 0.1--0.2 & 0.2--0.3 & 0.3--0.4 & 0.4--0.5 & 0.5--0.6 & 0.6--0.7 & 0.7--0.8 & 0.8--0.9 & 0.9--1.0 \\
        \midrule
0.0--0.1 & \phantom{$-$}1.00 & $-0.09$  & $-0.07$  & $-0.06$  & $-0.09$  & $-0.08$  & $-0.07$  & $-0.10$  & $-0.12$  & $-0.11$ \\
0.1--0.2 & &  \phantom{$-$}1.00 & \phantom{$-$}$0.04$  & \phantom{$-$}$0.03$  & \phantom{$-$}$0.01$  & $-0.17$  & $-0.16$  & $-0.22$  & $-0.12$  & $-0.10$ \\
0.2--0.3 & &  &  \phantom{$-$}1.00 & \phantom{$-$}$0.34$  & \phantom{$-$}$0.04$  & $-0.48$  & $-0.46$  & $-0.30$  & $-0.10$  & $-0.08$ \\
0.3--0.4 & &  &  &  \phantom{$-$}1.00 & \phantom{$-$}$0.03$  & $-0.59$  & $-0.63$  & $-0.24$  & $-0.08$  & $-0.07$ \\
0.4--0.5 & &  &  &  &  \phantom{$-$}1.00 & $-0.17$  & $-0.16$  & $-0.22$  & $-0.12$  & $-0.10$ \\
0.5--0.6 & &  &  &  &  &  \phantom{$-$}1.00 & \phantom{$-$}$0.52$  & \phantom{$-$}$0.14$  & $-0.06$  & $-0.07$ \\
0.6--0.7 & &  &  &  &  &  &  \phantom{$-$}1.00 & \phantom{$-$}$0.14$  & $-0.05$  & $-0.07$ \\
0.7--0.8 & &  &  &  &  &  &  &  \phantom{$-$}1.00 & $-0.07$  & $-0.09$ \\
0.8--0.9 & &  &  &  &  &  &  &  &  \phantom{$-$}1.00 & $-0.11$ \\
0.9--1.0 & &  &  &  &  &  &  &  &  &  \phantom{$-$}1.00\\
\bottomrule
      \end{tabular}
 % \end{center}
\end{table}

\clearpage
% Author List ----------------------------
%\input{LHCb_Authorship_flat_12-Dec-2016.tex}

\centerline{\large\bf LHCb collaboration}
\begin{flushleft}
\small
R.~Aaij$^{40}$,
B.~Adeva$^{39}$,
M.~Adinolfi$^{48}$,
Z.~Ajaltouni$^{5}$,
S.~Akar$^{59}$,
J.~Albrecht$^{10}$,
F.~Alessio$^{40}$,
M.~Alexander$^{53}$,
S.~Ali$^{43}$,
G.~Alkhazov$^{31}$
% ...

{\footnotesize \it
% ...
$ ^{5}$Clermont Universit{\'e}, Universit{\'e} Blaise Pascal, CNRS/IN2P3, LPC, Clermont-Ferrand, France\\
% ...
$ ^{10}$Fakult{\"a}t Physik, Technische Universit{\"a}t Dortmund, Dortmund, Germany\\
% ...
$ ^{31}$Petersburg Nuclear Physics Institute (PNPI), Gatchina, Russia\\
% ...
$ ^{39}$Universidad de Santiago de Compostela, Santiago de Compostela, Spain\\
$ ^{40}$European Organization for Nuclear Research (CERN), Geneva, Switzerland\\
% ...
$ ^{43}$Nikhef National Institute for Subatomic Physics, Amsterdam, The Netherlands\\
% ...
$ ^{48}$H.H. Wills Physics Laboratory, University of Bristol, Bristol, United Kingdom\\
% ...
$ ^{53}$School of Physics and Astronomy, University of Glasgow, Glasgow, United Kingdom\\
% ...
$ ^{59}$University of Cincinnati, Cincinnati, OH, United States\\
% ...
}
\end{flushleft}

\end{document}